\documentclass[english]{article}
\usepackage[T1]{fontenc}
\usepackage[latin9]{inputenc}
\usepackage[letterpaper]{geometry}
\usepackage{amsmath}
\usepackage{setspace}
\makeatletter

\providecommand{\tabularnewline}{\\}

\makeatother

\usepackage{babel}

\begin{document}

\title{\textbf{An algorithm to relate protein surface roughness with local
geometry of protein exterior shape.}}

\author{\textbf{Anirban Banerji}\\
\textbf{Bioinformatics Centre, University of Pune, Pune-411007,
Maharashtra, India}\\
\textbf{E-mail address: anirbanab@gmail.com}}
\maketitle
\begin{abstract}
\textbf{Changes in the extent of local concavity along with changes
in surface roughness of binding sites of proteins have long been considered
as useful markers to identify functional sites of proteins. However,
an algorithm that describes the connection between the simultaneous
changes of these important parameters - eludes the students of structural
biology. Here a simple yet general mathematical scheme is proposed
that attempts to achieve the same. Instead of n-dimensional random
vector description, protein surface roughness is described here as
a system of algebraic equations. Such description resulted in the
construction of a generalized index that not only describes the shape-change-vs-surface-roughness-change
process but also reduces the estimation error in local shape characterization.
Suitable algorithmic implementation of it in context-specific macromolecular
recognition can be attempted easily. Contemporary drug discovery studies
will be enormously benefited from this work because it is the first
algorithm that can estimate the change in protein surface roughness
as the local shape of the protein is changing (and vice-versa).}\\
\\

\end{abstract}
\textbf{\underbar{1: Introduction :}}\\
Proteins are not {}``classical solids''. Due to cytosolic Brownian
collisions and internal thermal fluctuations, shape of a protein suffers
from continuous fluctuations. As a result, the surface roughness undergoes
change also. Such small changes in shape and surface roughness of
a protein assume tremendous significance when it involves the (local)
shape of the binding site of an enzyme. Properties of protein binding
sites can broadly be classified in two, residue-specific properties
(amino acid type, surface accessibility, conservation, etc..) and
geometric properties (cavity shape, cavity depth, surface roughness,
etc ..). Relevant biophysical properties, viz. local electrostatics
and hydrophobicity - depend largely on the aforementioned properties.
Amongst all of them, changes in the extent of concavity of the active
sites of enzymes has long been recognized as a key feature to study
their structure-function relationship {[}1{]},{[}2{]}. Alongside this
recognition, came the acceptance that change in surface roughness
of patch of protein surface that envelops the local shape of binding
site, can serve as an equally useful marker while attempting to find
functional sites of that protein {[}3{]},{[}4{]}. While some previous
works could successfully quantify important aspects of geometry of
local shape on protein exterior {[}5{]}-{[}8{]}; an algorithm that
relates change in the concavity of the local shape to the change of
roughness of surface patch that holds this shape, could not be found.
Reason for this conspicuous absence can probably be attributed to
the complex nature of functional dependencies that influence surface
roughness of any given patch of protein surface. It is owing to these
(seemingly innumerable) dependencies that the surface roughness appears
to be an $n-$dimensional random variable. In this work, an attempt
is made to propose a simple yet rigorous mathematical procedure that
transforms the surface roughness description from that of an $n-$dimensional
random variable to a system of simple algebraic equations. Furthermore,
a generalized index is defined that captures the simultaneous changes
in local geometry and the roughness of the patch of protein surface
that envelops it. It can immediately be recognized that the algebraic
description of this problem is much easily implementable in computational
form, than the same attempted from the realm of $n-$dimensional continuous
random variables. Since the proposed algorithm is the very first one
(to the best of knowledge) in addressing the aforementioned problem,
it can be utilized in several applications in the extremely important
paradigm of objective description of protein flexibility.\\
\\
Proteins are not static rigid objects; their flexibility, especially
while binding to other molecules is a known fact{[}9{]}. Conformational
changes in enzymes upon substrate binding have been studied in details
{[}10, and references therein{]}. Questions regarding main-chain flexibility,
side-chain rotation and ligand flexibility were examined too{[}11,
12{]}. But a framework that attempts to combine change in protein
shape with change in its surface roughness, is difficult (if not impossible)
to find. The present model attempts to fill that void. Implementation
of this scheme in different contexts will help protein biophysics
study with concrete quantifiable markers. Contemporary studies in
the realm of drug discovery will be benefited from this algorithm
too, because it is (probably) for the first time that an attempt is
being made to connect the changing profiles of two most important
geometric properties of protein exterior.\\
\\
Since in-vivo macromolecular interactions ('docking' interactions
in particular) are time-dependent and context-dependent, the local
shape of the functional site, say the active site of enzyme, is expected
to undergo certain small yet significant change in its topography.
Such change in the local geometry of the shape necessarily corresponds
to a change in the extent of roughness of the surface that envelops
the local shape under consideration. Fractal dimension based markers
serve as reliable constructs to describe protein surface roughness
in objective terms. Hence to describe the entire situation, let $[s=0,\:1,\:2,\ldots,\: r]$
estimate the states of the local shape of the active site of an enzyme
when it is undergoing interaction with some inhibitor protein; and
let $[F=(f_{1},\: f_{2},\:\ldots,\: f_{n})]$ be the corresponding
vectors, describing the states representing surface roughness (expressed
with fractal dimensions) of the patch of the surface containing the
aforementioned local shape. We define a function $\phi\left(F\right)$,
on which the functional $Z=\Vert\phi\left(F\right)-s\Vert$ attains
a minimum. Since the magnitude of the functional $Z=\Vert\phi\left(F\right)-s\Vert$
has the capability to describe a change in either $\phi\left(F\right)$
or $s$; the minimum magnitude of the functional will imply a state
where the extent of change in $\phi\left(F\right)$ follows the trend
in extent of change in $\; s\;$, in the closest terms. Hence it is
this minimum magnitude of the functional $Z$ that is defined as generalized
index for studying the change in local shape of the enzyme active
site, because only with this magnitude of functional $Z$, the change
in both shape and surface roughness can be simultaneously described
in the best objective manner. One might expect that the minimum magnitude
of functional $Z$ will correlate positively with the most favorable
free-energetic cases. Correctness of such assertion will imply that
the patterns in observed free-energy profiles in thermodynamic studies
involving protein-ligand, enzyme-inhibitor, antigen-antibody interactions
etc., can reliably be studied with the algebraic framework proposed
here, without taking recourse to detailed simulation-based investigations.
This, in turn, might serve as a helpful tool while screening a large
number of compounds during high-throughput analyses, and also, while
attempting to characterize and compare between classes of macromolecular
interaction modes.\\
\\
\\
\\
\\
\textbf{\underbar{2: Methodology :}}\\
\textbf{2.1: Theoretical framework:}\\
It is clear that the solution of the given problem is determined
by the choice of the norm $\Vert.\Vert$ in the context of calculation
of functional $Z$. \\
\\
To describe the situation from an unbiased perspective, one may
start by considering $F$ as $n-$dimensional continuous random variable
with correlation matrix $C$ associated with it. (A correlation matrix
describes correlation among $\, M\;$ variables. It is a square symmetrical
$\; M\,\times\, M\;$ matrix with the $(ij)^{th}\;$ element equal
to the correlation coefficient $r_{ij}\;$ between the $(i)^{th}\;$
and the $(j)^{th}\;$ variable. The diagonal elements (correlations
of variables with themselves) are always equal to unity. Furthermore,
correlation matrix is always symmetric because correlation between
$\; X_{i}\;$ and $\; X_{j}\;$ is the same as the correlation between
$\; X_{j}\;$ and $\; X_{i}\;$.). One may then define $\phi\left(F\right)$
as a predictor of a random variable $s$ having a minimum mean square
value. Thus the functional $Z$ can be re-written as :\\
\\
\begin{equation}
Z=\sum_{s=0}^{r}p_{s}E\left[\left(\phi\left(F\right)-s\right)^{2}|\; s\right]\end{equation}
\\
where $E$ represents the expectation operator and $p_{s}$ denotes
the probability of occurrence of any arbitrarily chosen state of the
shape, $s$. In the case where $E\left(\phi\left(F\right)|s\right)=s$
for all $[s=0,1,\ldots,r]$, - the given functional state attains
a minimum.\\
\\
To describe the situation from a bottom-up perspective, i.e.,
describing with respect to an individual fractal dimension $(f_{k})$
that represents a particular value of surface roughness, one will
have : $E\left(f_{k}|s\right)=s$ for all $s$ and $[k=0,1,\ldots,r].$
However, here we assume that the form of the relations $E(f_{i}|s)$
and the coefficients $\alpha_{i}$, $(i=1,2,...,n)$ remain unchanged
in time.\\
\\
It is then possible to represent $\phi\left(F\right)$ as a linear
combination of indices :\\
\\
\begin{equation}
\phi\left(F\right)=\sum_{i=1}^{n}\alpha_{i}f_{i}\end{equation}
where $\sum_{i=1}^{n}\alpha_{i}=1$; since in this case $E\left(\phi\left(F\right)|s\right)=s$.
The coefficients $\alpha_{1},\alpha_{2},\ldots,\alpha_{n}$ are evaluated
from the minimum condition of the Lagrangian :\\
\\
\begin{equation}
\widetilde{Z}\left(\alpha,\lambda\right)=\sum_{s}p_{s}E\left[\left(\sum_{i=1}^{n}\alpha_{i}\left(f_{i}-s\right)\right)^{2}|s\right]+\lambda\left(\sum_{i=1}^{n}\alpha_{i}-1\right)\end{equation}
\\
Differentiating $\widetilde{Z}\left(\alpha,\lambda\right)$ with
respect to $\alpha_{i}$ $(i=1,2,...,n)$ and equating the derivatives
to zero one obtains a system of algebraic equations in $\alpha$ and
$\lambda$ :\\
\\
\begin{equation}
\sum_{i=1}^{n}\alpha_{i}\sum_{s}p_{s}E\left[\left(f_{k}-s\right)\left(f_{i}-s\right)|s\right]+\lambda=0\end{equation}
\\
where $(k=1,2,...,n)$ and $\sum\alpha_{i}=1$.\\
\\
In case where $C$ is the diagonal matrix, the solution of the
given system can be written as :\\
\begin{equation}
\alpha_{i}=\left(\sum_{k=1}^{n}\prod_{j\neq k}C_{jj}\right)^{-1}\prod_{k\neq i}C_{kk}\end{equation}
\\
and\\
\begin{equation}
\lambda=-\left(\sum_{k=1}^{n}\prod_{j\neq k}C_{jj}\right)^{-1}\prod_{k=i}C_{kk}\end{equation}
\\
The variance of the generalized index will be given by:\\
\begin{equation}
E\left(\phi\left(F\right)-s\right)^{2}=\left(\sum_{k=1}^{n}\frac{1}{C_{kk}}\right)^{-1}\end{equation}
\\
Since C$_{kk}$, $(k=1,2,...,n)$, are positive and bounded, we
arrived at : \\
\begin{equation}
\underset{n\rightarrow\infty}{lim}E\left(\phi\left(F\right)-s\right)^{2}=0.\end{equation}
Eq$^{n}$-8 shows unambiguously how the introduction of the generalized
index can reduce the estimation error in local shape characterization
and how such an index makes it possible to transform the problem from
the complex analysis of an $n-$dimensional variable to that of involving
scaler parameters. This particular aspect of the present work makes
it easily compatible with the computational tool-kit. It is achieved
without compromising with the necessary rigor that is essential to
describe the intricate (time-dependent and context-dependent) coupling
between simultaneous changes of two variables with large dependencies.\\
\\
In some special cases, the regressions $E\left(f_{k}|s\right)=\psi_{k}\left(s\right)$
, (k = 1, 2, ..., n) can be non-linear. These can be reduced by transforming
regressive relations to a linear form; for example, with the help
of the polynomials $\mu_{k}\left(f_{k}\right)$ , $(k=1,2,...,n)$;
so that one obtains $E\left[\mu_{k}\left(f_{k}\right)|s\right]=s$.
Then the generalized index will be linear combination of polynomials
:\\
\\
$\phi(F)=\sum_{i=1}^{n}\alpha_{i}\mu_{i}(f_{i})$.\\
\\
The proposed methodology is mathematical and general in nature.
Outlines of how it can be transformed to the realm of algorithm is
provided below.\\
\textbf{\underbar{}}\\
\\
\\
\textbf{\underbar{2.2: Algorithmic Implementation :}}\\
\textbf{2.2.1: Algorithm to construct correlation matrix :}\\
To begin with, cavity coordinates from a set of (say, 10) different
protein structures are taken.\\
\\
\textbf{Step 1):} For every cavity from distinct proteins, the
FD, concavity, planarity and overall surface area - are calculated,
during the entire interval of interaction.\\
\\
\textbf{Step 2) :} As a result of step-1 operations, for 10 such
distinct cavities, we will obtain a $\;10\,\times\,4\;$ matrix (in
general, for 'n' such cavities, one would obtain a $\; n\,\times\,4\;$
matrix). Here, FD will be placed in the 1st column, followed by concavity,
planarity, overall surface area in column 2, 3, 4. These columns are
named accordingly as X1, X2, X3 and X4.\\
\\
\textbf{Step 3) :} For the column X1 (comprised of instances of
X1, viz. x\_1, x\_2, ...),\\
$\qquad\qquad$3.1) the mean of x\_1, x\_2, ... etc. - is calculated.\\
$\qquad\qquad$3.2) column X1 is populated with new entries; it
is populated by entries x\_1\_centered, x\_2\_centered etc.; where
these replacement entries are defined as:\\
x\_1\_centered = (x\_1-x\_1\_mean), x\_2\_centered is defined
by by x\_2=(x\_2-x\_1\_mean), etc. From now on these will be denoted
as $(x_{1}centered)$, $(x_{2}centered)$ etc. The column therefore
should be renamed as X1\_centered. \\
$\qquad\qquad$3.3) Repeat the same, on the other columns too.
A matrix with these columns will be called as (X\_centered).\\
\\
\textbf{Step 4) :} Construct the transpose of this matrix, (X\_centered)'
; that is, a matrix with its rows being X1\_centered, X2\_centered,
etc..\\
\\
\textbf{Step 5) :} Calculate matrix multiplication between (X\_centered)
and (X\_centered)'. \\
- This will give $\sum(x_{1}centered).(x_{2}centered)$. Thus,
for 4 variables, as enlisted in Step-2, 4x4 matrices will be generated;
where the terms will be :\\
 \begin{tabular}{|c|c|c|c|}
\hline 
$\sum(x_{1}centered)^{2}$ &  &  & \tabularnewline
\hline
\hline 
$\sum(x_{1}centered).(x_{2}centered)$ & $\sum(x_{2}centered)^{2}$ &  & \tabularnewline
\hline 
$\sum(x_{1}centered).(x_{3}centered)$ & $\sum(x_{2}centered).(x_{3}centered)$ & $\sum(x_{3}centered)^{2}$ & \tabularnewline
\hline 
$\sum(x_{1}centered).(x_{4}centered)$ & $\sum(x_{2}centered).(x_{4}centered)$ & $\sum(x_{3}centered).(x_{4}centered)$ & $\sum(x_{4}centered)^{2}$\tabularnewline
\hline
\end{tabular}\\
\\
\textbf{Step 6) :} Calculate the standard-deviation for 4 cases,
viz. standard-deviation for X1\_centered, X2\_centered, etc.\\
\\
\textbf{Step 7) :} Construct the following matrix :\\
\begin{tabular}{|c|c|c|c|}
\hline 
$\frac{\sum(x_{1}centered)^{2}}{N-1}$ &  &  & \tabularnewline
\hline
\hline 
$\frac{\sum(x_{1}centrd).(x_{2}centrd)}{N-1}$ & $\frac{\sum(x_{2}centered)^{2}}{N-1}$ &  & \tabularnewline
\hline 
$\frac{\sum(x_{1}centrd).(x_{3}centrd)}{N-1}$ & $\frac{\sum(x_{2}centrd).(x_{3}centrd)}{N-1}$ & $\frac{\sum(x_{3}centered)^{2}}{N-1}$ & \tabularnewline
\hline 
$\frac{\sum(x_{1}centrd).(x_{4}centrd)}{N-1}$ & $\frac{\sum(x_{2}centrd).(x_{4}centrd)}{N-1}$ & $\frac{\sum(x_{3}centrd).(x_{4}centrd)}{N-1}$ & $\frac{\sum(x_{4}centered)^{2}}{N-1}$\tabularnewline
\hline
\end{tabular}\\
\\
Since, there can be countably infinite number of protein surface
patches for protein cavity vibrations, and since not all of them are
being subjected to investigation, this work will be involving samples
and not the population. Hence the dividing term will be (N-1), instead
of N.\\
\\
\\
\textbf{Step 8) :} Construct the \textbf{Correlation matrix},
given by :\\
\begin{tabular}{|c|c|c|c|}
\hline 
$\frac{\sum(x_{1}centered)^{2}}{\sigma_{1}\sigma_{1}(N-1)}$ &  &  & \tabularnewline
\hline
\hline 
$\frac{\sum(x_{1}centrd).(x_{2}centrd)}{\sigma_{1}\sigma_{2}(N-1)}$ & $\frac{\sum(x_{2}centered)^{2}}{\sigma_{2}\sigma_{2}(N-1)}$ &  & \multicolumn{1}{c|}{}\tabularnewline
\hline 
$\frac{\sum(x_{1}centrd).(x_{3}centrd)}{\sigma_{1}\sigma_{3}(N-1)}$ & \multicolumn{1}{c|}{$\frac{\sum(x_{2}centrd).(x_{3}centrd)}{\sigma_{2}\sigma_{3}(N-1)}$} & $\frac{\sum(x_{3}centered)^{2}}{\sigma_{3}\sigma_{3}(N-1)}$ & \tabularnewline
\hline 
$\frac{\sum(x_{1}centrd).(x_{4}centrd)}{\sigma_{1}\sigma_{4}(N-1)}$ & $\frac{\sum(x_{2}centrd).(x_{4}centrd)}{\sigma_{2}\sigma_{4}(N-1)}$ & $\frac{\sum(x_{3}centrd).(x_{4}centrd)}{\sigma_{3}\sigma_{4}(N-1)}$ & $\frac{\sum(x_{4}centered)^{2}}{\sigma_{4}\sigma_{4}(N-1)}$\tabularnewline
\hline
\end{tabular}\\
\\
\\
\textbf{2.2.2: Calculation of $\alpha_{i}$(coefficients to satisfy
Lagrangian minima) and $\phi\left(F\right)$(the generalized index)
:}\\
In the 4x4 correlation matrix (given above), n=4; k=1,2,3,4; j=1,2,3,4;
and i=4. The variable 'n' describes the number of rows; k and j are
counters that vary from 1 to 4. Variable i decides the number of coefficients
needed to obtain the Lagrangian minima. This variable can be chosen
arbitrarily; in the present case it has been chosen to be 4.\\
\\
\textbf{Step-1) :} We zoom in on equation-5.\\
\\
The following calculations are made : \begin{multline*}
denom=\left[\sum(x_{2}centrd)^{2}/\sigma_{2}\sigma_{2}(N-1)\times\sum(x_{3}centrd)^{2}/\sigma_{3}\sigma_{3}(N-1)\times\sum(x_{4}centrd)^{2}/\sigma_{4}\sigma_{4}(N-1)\right]+\\
\left[\sum(x_{1}centrd)^{2}/\sigma_{1}\sigma_{1}(N-1)\times\sum(x_{3}centrd)^{2}/\sigma_{3}\sigma_{3}(N-1)\times\sum(x_{4}centrd)^{2}/\sigma_{4}\sigma_{4}(N-1)\right]+\\
\left[\sum(x_{1}centrd)^{2}/\sigma_{1}\sigma_{1}(N-1)\times\sum(x_{2}centrd)^{2}/\sigma_{2}\sigma_{2}(N-1)\times\sum(x_{4}centrd)^{2}/\sigma_{4}\sigma_{4}(N-1)\right]+\\
\left[\sum(x_{1}centrd)^{2}/\sigma_{1}\sigma_{1}(N-1)\times\sum(x_{2}centrd)^{2}/\sigma_{2}\sigma_{2}(N-1)\times\sum(x_{3}centrd)^{2}/\sigma_{3}\sigma_{3}(N-1)\right]\end{multline*}
\\
\begin{multline*}
Numerator_{1}=\left[\sum(x_{2}centrd)^{2}/\sigma_{2}\sigma_{2}(N-1)\times\sum(x_{3}centrd)^{2}/\sigma_{3}\sigma_{3}(N-1)\times\sum(x_{4}centrd)^{2}/\sigma_{4}\sigma_{4}(N-1)\right]\end{multline*}
\\
\begin{multline*}
Numerator_{2}=\left[\sum(x_{1}centrd)^{2}/\sigma_{1}\sigma_{1}(N-1)\times\sum(x_{3}centrd)^{2}/\sigma_{3}\sigma_{3}(N-1)\times\sum(x_{4}centrd)^{2}/\sigma_{4}\sigma_{4}(N-1)\right]\end{multline*}
\\
\begin{multline*}
Numerator_{3}=\left[\sum(x_{1}centrd)^{2}/\sigma_{1}\sigma_{1}(N-1)\times\sum(x_{2}centrd)^{2}/\sigma_{2}\sigma_{2}(N-1)\times\sum(x_{4}centrd)^{2}/\sigma_{4}\sigma_{4}(N-1)\right]\end{multline*}
\\
\begin{multline*}
Numerator_{4}=\left[\sum(x_{1}centrd)^{2}/\sigma_{1}\sigma_{1}(N-1)\times\sum(x_{2}centrd)^{2}/\sigma_{2}\sigma_{2}(N-1)\times\sum(x_{3}centrd)^{2}/\sigma_{3}\sigma_{3}(N-1)\right]\end{multline*}
\\
Finally, we assign $\alpha_{1}=Numerator_{1}/denom$, $\alpha_{2}=Numerator_{2}/denom$,
$\alpha_{3}=Numerator_{3}/denom$, $\alpha_{4}=Numerator_{4}/denom$\\
\\
\\
\textbf{Step-2) :} These findings are applied to equation-2.\\
We can choose either of the methods. In the \textbf{\underbar{first}},
the most prominently different 4 FD values ($f_{i}$) are chosen out
of 10 FDs obtained from as many cavity structures. These are sorted.
Then $\phi\left(F\right)=\sum_{i=1}^{n}\alpha_{i}f_{i}$ is calculated
for $f_{i}$ with ascending order and for descending order. The mean
of $\phi\left(F\right)$ from these two (asending and descending order
$f_{i}$ ), will assume the final magnitude of $\phi\left(F\right)$.\\
Or else,\\
the \textbf{\underbar{second}} method can be implemented. Here
4 $f_{i}$ values are chosen randomly, from the pool of obtained $f_{i}$s.
Then an arbitrary number of $\phi\left(F\right)$ are generated. A
mean of these $\phi\left(F\right)$s, will provide the final $\phi\left(F\right)$.\\
\\
\\
\textbf{Step-3) :} Following the calculation of $\phi\left(F\right)$,
the functional $Z=\Vert\phi\left(F\right)-s\Vert$ is calculated.\\
\\
The minimum magnitude of the functional $Z=\Vert\phi\left(F\right)-s\Vert$
corresponds to a state where the extent of change in $\phi\left(F\right)$
follows the trend in extent of change in $\; s\;$, in the closest
terms. Magnitude of the functional $Z$ , therefore, quantifies the
pattern obtained from analyzing the features of functional site geometries
of the chosen set of proteins. In other words, these magnitudes can
be compared with profiles of free-energies, typically reported for
various known types of macromolecular interactions.\\
\\
\textbf{\underbar{3: Conclusion :}}\\
It is easy to observe that the scope of applications of the proposed
methodology is not merely restricted to protein active site studies;
due to its generalized approach, it might be helpful in several problems
in the field of protein flexibility study. Researchers in the fields
like protein-DNA interactions and protein-small molecule interactions,
which involve large number of quantifiable parameters to describe
(mutual) flexibility of the biomolecules, will probably find the present
methodology to be helpful. However, simple and useful as this technique
is, it is not immune to limitations. For example, the algorithm proposed
in the current work might not be efficient when applied to cases where
functional sites are not appreciably concave. For example, since many
of the protein-protein interaction interfaces are known to be planer
{[}13{]}, the scope of the application of the present algorithm might
be restricted on such cases. Having said that, since for many proteins,
the functional sites are known to be containing concave shapes (in
the form of clefts and pockets) {[}5{]}, the spectrum of contexts
where the present algorithm can be applied assumes an impressive range.
The form of simple algebraic equations with which the final description
of this complex biophysical interaction is expressed, makes it easily
implementable with elementary computational apparatus.\\
\\
\textbf{\underbar{References :}}\\
{[}1{]} R.A. Lewis, Clefts and binding sites in protein receptors,
Meth. Enzymol. 202 (1991) 126\textendash{}156.\\
{[}2{]} F.K. Pettit, E. Bare, A. Tsai, J.U. Bowie, HotPatch: A
Statistical approach to finding biologically relevant features on
protein surfaces. J. Mol. Biol. 369 (2007) 863\textendash{}879.\\
{[}3{]} M. Lewis, D.C. Rees, Fractal surfaces of proteins, Science,
230 (1985) 1163-1165.\\
{[}4{]} F.K. Pettit, J.U. Bowie, Protein surface roughness and
small molecular binding sites, J. Mol. Biol. 285 (1999) 1377\textendash{}1382.\\
{[}5{]} R.A. Laskowski, N.M. Luscombe, M.B. Swindells, J.M. Thornton,
Protein clefts in molecular recognition and function. Prot. Sci. 5
(1996) 2438\textendash{}2452.\\
{[}6{]} K.P. Peters, J. Fauck, C. Frommel, The automatic search
for ligand binding sites in proteins of known three-dimensional structure
using only geometric criteria, J. Mol. Biol. 256 (1996) 201\textendash{}213.\\
{[}7{]} J. Liang, H. Edelsbrunner, C. Woodward, Anatomy of protein
pockets and cavities: measurement of binding site geometry and implications
for ligand design, Prot. Sci. 7 (1998) 1884\textendash{}1897.\\
{[}8{]} G.P. Brady Jr., P.F. Stouten, Fast prediction and visualization
of protein binding pockets with PASS, J. Comp-Aided Mol. Des. 14 (2000)
383\textendash{}401.\\
{[}9{]} R. Huber, W.S. Bennett Jr., Functional significance of
flexibility in proteins. Biopolymers 22 (1983) 261-279.\\
{[}10{]} A. Gutteridge, J.M. Thornton, Conformational changes
observed in enzyme crystal structures upon substrate binding, J. Mol.
Biol. 346 (2005) 21-28.\\
{[}11{]} M. I. Zavodszky, M. Lei, A. R. Day, M. F. Thorpe, L.
A. Kuhn, Modeling Correlated Main-chain Motions in Proteins for Flexible
Molecular Recognition, Prot.: Struct. Funct. Bioinf. 57 (2004) 243-261.\\
{[}12{]} M.I. Zavodszky, L.A. Kuhn, Side-Chain Flexibility in
Protein-Ligand Binding: The Minimal Rotation Hypothesis, Protein Sci.
14 (2005) 1104-1114.\\
{[}13{]} S. Jones, J.M. Thornton, Prediction of protein-protein
interaction sites using patch analysis, J. Mol. Biol. 272 (1997) 133\textendash{}143.\\
\\
\\

\end{document}